# Motion-corrected and high-resolution anatomically-assisted (MOCHA) reconstruction of arterial spin labelling MRI

Abolfazl Mehranian[1], Colm J. McGinnity[2], Radhouene Neji[1,3], Claudia Prieto[1], Alexander Hammers[2], Enrico De Vita[1†] and Andrew J. Reader[1†]

School of Biomedical Engineering and Imaging Sciences, King's College London, [1]Department of Biomedical Engineering and [2]King’s College London & Guy’s and St Thomas’ PET Centre, St. Thomas' Hospital, London, UK, [3]MR Research Collaborations, Siemens Healthcare Limited, Frimley, UK

[†]These authors contributed equally to this work

**Corresponding author**:
Enrico De Vita, PhD.
School of Biomedical Engineering and Imaging Sciences
King's College London
3rd Floor, Lambeth Wing
St Thomas' Hospital
London, SE1 7EH
Email: Enrico.Devita@kcl.ac.uk

**Running head title:**
High-resolution ASL image reconstruction

**Word count**: ~6000

# Abstract

**Purpose**: A model-based reconstruction framework is proposed for MOtion-Corrected and High-resolution anatomically-Assisted (MOCHA) reconstruction of ASL data. In this framework, all low-resolution ASL control-label pairs are used to reconstruct a single high-resolution cerebral blood flow (CBF) map, corrected for rigid motion, point-spread-function (PSF) blurring and partial-volume effect (PVE).

**Methods**: Six volunteers were recruited for CBF imaging using PCASL labelling, 2-shot 3D-GRASE sequences and high-resolution T1-weighted MRI. For two volunteers, high-resolution scans with double and triple resolution in the partition direction were additionally collected. Simulations were designed for evaluations against a high-resolution ground-truth CBF map, including a simulated hyper-perfused lesion and hyper/hypo-perfusion abnormalities. MOCHA was compared to standard reconstruction and a 3D linear regression (3DLR) PVE correction method and was further evaluated for acquisitions with reduced control-label pairs and k-space undersampling.

**Results**:

MOCHA reconstructions of low-resolution ASL data showed enhanced image quality particularly in the partition direction. In simulations, both MOCHA and 3DLR provided more accurate CBF maps than the standard reconstruction, however MOCHA resulted in the lowest errors and well delineated the abnormalities. MOCHA reconstruction of standard-resolution *in-vivo* data showed good agreement with higher-resolution scans requiring 4× and 9× longer acquisitions. MOCHA was found to be robust for 4×-accelerated ASL acquisitions, achieved by reduced control-label pairs or k-space undersampling.

**Conclusion**: MOCHA reconstruction reduces PVE by direct reconstruction of CBF maps in the high-resolution space of the corresponding anatomical image, incorporating motion correction and PSF modelling. Following further evaluation, MOCHA should promote the clinical application of ASL.

# 1. Introduction

Arterial spin labelling (ASL) is a non-invasive perfusion-weighted magnetic resonance imaging (MRI) technique for quantification of cerebral blood flow (CBF) (1), using magnetically labelled blood water as an endogenous contrast agent. In this technique, blood spins are typically labelled by inversion before flowing into the imaging volume, with pseudo-continuous ASL (PCASL) currently as the preferred method (1). The difference between label and control (i.e. non-labelled) images produces a signal proportional to the local tissue blood flow (2). ASL has an intrinsically low signal to noise ratio (SNR), as the volume of labelled blood is only ~1-2% of total cerebral blood volume (~4-5%) and the fact that the magnetic label decays by the T1 relaxation time of blood while it flows from labelling region to imaging volume. To allow the labelled blood to reach the imaging volume, the ASL signal is acquired following a post-label delay (PLD) time. Short PLDs are associated with less T1 decay and higher SNR, however too short PLDs may be insufficient for full arrival of labelled blood into the tissues leading to inaccurate CBF quantification.

To improve SNR, typically 10-50 control-label (C-L) pairs with low nominal spatial resolution (in-plane: 3-4 mm, through-plane: 4-8 mm) are acquired and averaged (1). In addition, background suppression (3), 3D readout sequences (4) and parallel imaging (5) are also used to respectively suppress static tissues, increase the SNR and brain coverage, and reduce acquisition time. While reducing spatial resolution improves SNR, it results in partial volume averaging of grey (GM) and white matter (WM) CBF (6). Moreover, the widely-used 3D readout sequences such as gradient and spin echo (GRASE) (7) can introduce substantial through-plane blurring (due to the T2 decay of signal across echo trains) and hence contribute to partial volume effects (PVE). For single-shot GRASE, the through-plane point spread function (PSF) has been reported to be from 1.5 to 1.9 voxels full width at half maximum (FWHM) (8). Segmented acquisition schemes help minimise this effect, however as the number of shots increases, acquisition time and sensitivity to motion also increase (9).

For partial volume correction (PVC), existing methods aim to unmix GM and WM signals (overlapping in low-resolution acquisitions) using partial volume (PV) estimates obtained from anatomical MR images. They are linear regression (LR) (10) , modified least trimmed squares (11) or Bayesian inference for ASL (6). PV estimation requires accurate registration, segmentation and downsampling of the anatomical MR images into the ASL image resolution, which are prone to errors (12). These PVC methods can be preceded by a deconvolution pre-processing step to reduce the PSF blurring (9) however, deconvolution is known to amplify noise and can result in Gibbs ringing artefacts. PVE can be reduced by increasing the acquisition's spatial resolution, however the reduced SNR, requires more averaging, i.e. longer acquisition time that increase motion sensitivity. Hence, a number of denoising (13) and undersampled MRI techniques (14) have been proposed to reduce noise while using as few averages as possible. Currently, PVC involves several pre-processing steps of ASL images (deconvolution, denoising and motion correction) and of structural MR image (registration, segmentation and downsampling) (15). The actual PVC step is then typically carried out in the image space of the low-resolution C-L pairs, while operating at a higher resolution might improve their performance (16).

In this study, we propose a framework for reconstruction of low-resolution ASL data into the high-resolution space of the anatomical images, corrected for motion, PSF blurring and under-sampling artefacts, with additional noise reduction. To effectively reduce noise and PVE, firstly, all C-L pairs are simultaneously used to reconstruct a single perfusion-weighted ASL image compared to the standard methods in which the C-L data are separately reconstructed, motion corrected, subtracted and then averaged. Secondly, a smoothness prior, weighted by the anatomical image, is utilized to assist the reconstruction of the target high-resolution perfusion image. In this work, the proposed Motion-Corrected and High-resolution anatomically-Assisted (MOCHA) ASL image reconstruction method was evaluated using simulations and *in vivo* datasets and compared with the standard reconstruction methods and a 3-dimensional LR (3DLR) method (17).

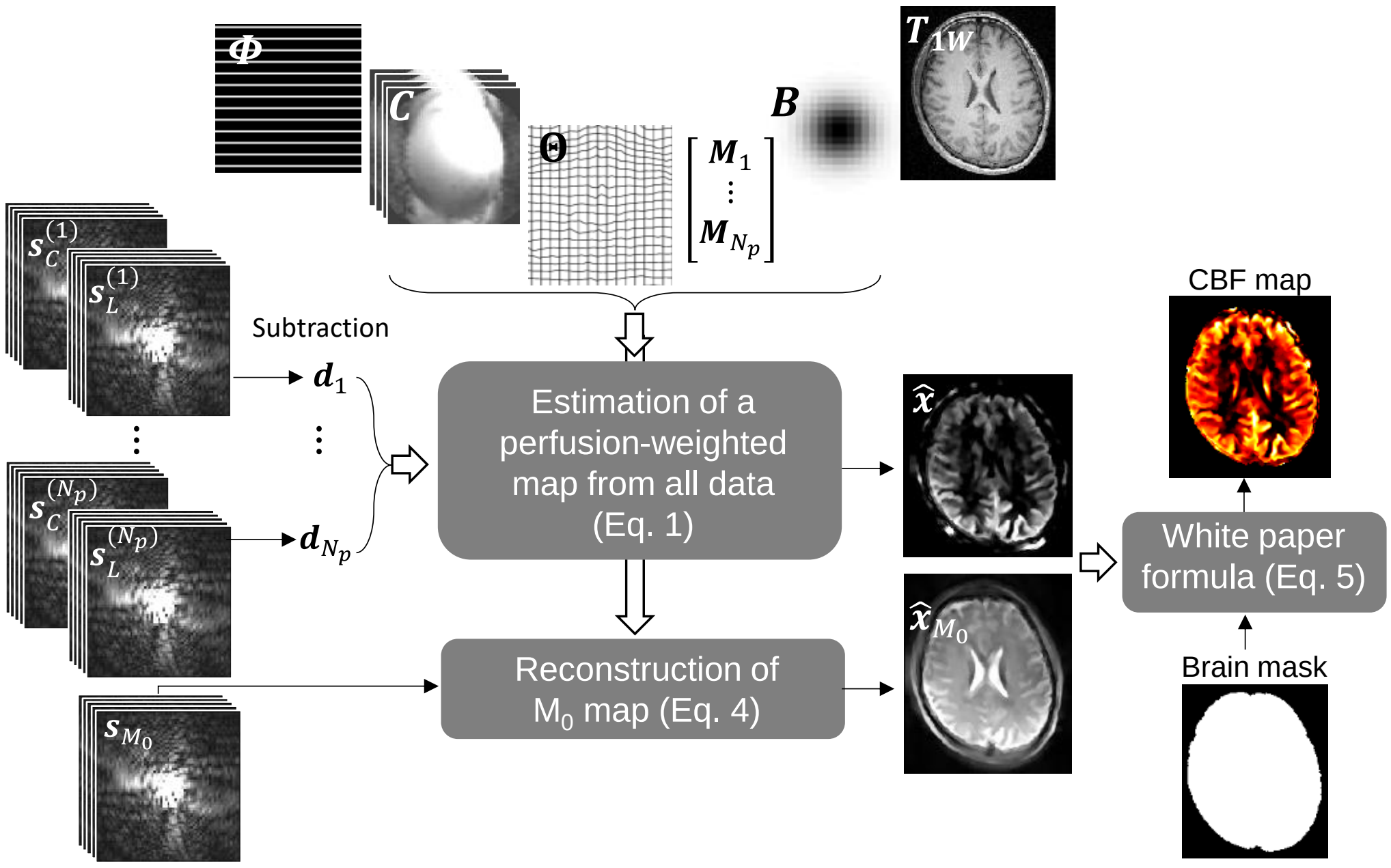

**Figure 1.** Flowchart of the MOCHA reconstruction algorithm.

# 2. Methods

## 2.1. MOCHA reconstruction

Reconstruction of a high-resolution perfusion-weighted image, $\boldsymbol{x} \in \mathbb{C}^{N_h}$, from $N_p$ pairs of low-resolution C-L ASL data was formulated as the following model-based minimization problem (18):

$$\hat{\boldsymbol{x}} = \underset{\boldsymbol{x}}{\operatorname{argmin}} \left\{ \frac{1}{2N_p} \sum_{i}^{N_p} \|\boldsymbol{E}\boldsymbol{T}_i\boldsymbol{B}\boldsymbol{x} - \boldsymbol{d}_i\|_{\boldsymbol{W}}^2 + \beta R(\boldsymbol{x}) \right\} \qquad [1]$$

where $\boldsymbol{d}_i \in \mathbb{C}^{N_m L}$ is the element-wise subtraction of the $i^{\text{th}}$ control and label multi-channel *k*-space data (namely, perfusion-weighted data), $N_m$ and $L$ are the number of *k*-space samples and the number of coils. $\boldsymbol{B} \in \mathbb{R}^{N_h \times N_h}$ is a convolution operator used to model PSF blurring of the MR sequence in image space, where $N_h$ is the number of voxels in the high-resolution MR image. $\boldsymbol{T}_i = \boldsymbol{\Theta}\boldsymbol{D}\boldsymbol{M}_i \in \mathbb{R}^{N_l \times N_h}$ is composed of the rigid transformation of $\boldsymbol{x}$ to the $i^{\text{th}}$ motion state ($\boldsymbol{M}_i$) (see *section 2.3.1*), downsampling ($\boldsymbol{D}$) to ASL low-resolution space and non-rigid geometric distortion ($\boldsymbol{\Theta}$) induced by $B_0$ field inhomogeneity, which was set to identity in this work. $N_l$ is the number of voxels in the ASL space. $\boldsymbol{E} = (\boldsymbol{I}_L \otimes \boldsymbol{\Phi}\boldsymbol{F})\boldsymbol{C} \in \mathbb{C}^{N_m L \times N_l}$ is composed of coil sensitivity matrix of $L$ coils ($\boldsymbol{C} \in \mathbb{C}^{N_l L \times N_l}$), Fourier transform ($\boldsymbol{F} \in \mathbb{C}^{N_l \times N_l}$) and *k*-space undersampling matrix ($\boldsymbol{\Phi} \in \mathbb{R}^{N_m \times N_l}$), with $N_m \leq N_l$ samples and $\otimes$ is the Kronecker product and $\boldsymbol{I}_L$ an identity matrix of size $L$. $\boldsymbol{W} \in \mathbb{R}^{N_m L \times N_m L}$ is the weighting matrix obtained from the inversion of the noise covariance matrix (19), which was set to identity in this work. Figure 1 illustrates a flow-chart describing the forward model used in Eq. [1]. $R(\boldsymbol{x})$ is a penalty function defined as a weighted quadratic prior as follows:

$$R(\boldsymbol{x}) = \sum_{j}^{N_h} \sum_{b \in \mathcal{N}_j} \omega_{jb} \xi_{jb} \left(x_j - x_b\right)^2 \quad [2]$$

which aims to suppress noise and artefacts based on the intensity differences between voxels $j$ and $b$ in the neighbourhood $\mathcal{N}_j$, while preserving boundaries using the similarity coefficients $\omega_{jb}$, calculated from the anatomical image. $\xi_{jb}$ are proximity coefficients used to modulate the intensity differences based on their Euclidian distance. The $\beta$ in Eq. [1] is a regularization parameter. In this study, the similarity coefficients were defined using Gaussian kernels (20) as follows:

$$\omega_{jb} = \frac{1}{\sqrt{2\pi}\sigma} \exp\left(-\frac{\left(v_j - v_b\right)^2}{2\sigma^2}\right) \quad [3]$$

where $\boldsymbol{v} \in \mathbb{R}^{N_h}$ is the MR anatomical image, $\sigma$ is a shape hyper-parameter. The reconstruction method in Eq. [1] aims to perform PVC using a higher-resolution image grid and PSF modelling. As described in *Section 2.3.2*, for CBF quantification, a calibration image ($M_0$) is also acquired during ASL scan. $M_0$ images were reconstructed using a method similar to Eq. [1] but devised for individual *k*-spaces as follows:

$$\hat{\boldsymbol{x}}_{M_0} = \underset{x_{M_0}}{\operatorname{argmin}} \left\{ \frac{1}{2} \left\| \boldsymbol{E}\boldsymbol{T}_{M_0} \boldsymbol{B} \boldsymbol{x}_{M_0} - \boldsymbol{s}_{M_0} \right\|_W^2 + \beta R\left(\boldsymbol{x}_{M_0}\right) \right\} \quad [4]$$

where $\boldsymbol{s}_{M_0}$ is the k-space data of $M_0$ dataset and $R$ is the same as defined in Eq. [2]. In this work, Eqs. [1, 4] were solved using the steepest decent (SD) algorithm (see Appendix A).

## 2.2. *In vivo* data acquisition

Six healthy volunteers (all males, mean age (±sd) 40±5 years) were scanned on a Siemens 3T Biograph PET-MR scanner with a 12-channel head coil. For perfusion imaging, a PCASL labelling scheme (21) was used with a centre-out 3D-GRASE readout with the following parameters: repetition time (TR): 4000 ms, echo time (TE): 17.62 ms, flip angle (FA): 150° (chosen to reduce blurring in the partition direction), image matrix: 64×62×29, nominal resolution: 4×4×4mm$^3$, reconstruction field of view (FOV): 256×256×104 mm$^3$, slice oversampling: 10%, turbo factor: 29, echo planar imaging (EPI) factor: 31, number of shots (segments): 2, bandwidth: 3126 Hz/pixel, background suppression (BS): on, labelling duration: 1500 ms, PLD: 1800 ms, number of C-L pairs: 20, acquisition time (TA): 5 min 40 sec. After excitation pulse, a 3-line reference scan was acquired without phase encoding blips for phase correction. For CBF quantification, a calibration scan was performed using the same readout but without labelling and BS (*see section 2.3.2*). For BS, a pre-saturation was applied before the PCASL train and then 2 global inversion pulses during PLD, with positions chosen to minimise signal for tissues with T1s between 700 and 1400ms. For two participants, high-resolution ASL scans with double and triple resolution in the partition direction (i.e. 2.0 and 1.33 mm) were additionally acquired (parameters remained the same except for the use of 4 and 8 shots, doubled and tripled number of repeats to match the SNR of the lower-resolution acquisition, resulting in 22 min and 48 min 52 s scans, respectively). A magnetization prepared rapid gradient echo (MPRAGE) sequence was acquired with TR/TE/TI: 1700/2.63/900 ms, FA: 9°, FOV: 236×270×194 mm$^3$, resolution: 1.05×1.05×1.1 mm$^3$, image matrix: 224×256×176 and TA: 6 min 20 sec. This study was approved by the research ethics committee of our institution and written informed consent was obtained from all participants.

## 2.3. Data preprocessing

*2.3.1. Motion estimation*

To estimate head motion during acquisition, the C-L image pairs are individually reconstructed in their native resolution and processed with SPM12 (22) and the ASL-toolbox (15). For this purpose, the $M_0$ image of each ASL dataset is registered to its corresponding T1-weighted (T1w) MR image using SPM with default co-registration parameters. The ASL-toolbox rigidly registers all control and label images to the calibration scan whilst regressing out the potential registration errors caused by the intensity differences of C-L images (23). Finally, the estimated transformations are used for motion correction. As MOCHA relies on perfusion-weighted data (i.e. subtraction of control and label *k*-spaces), the motion between control and label data within a pair was neglected, while motion between pairs was estimated and corrected. For the other reconstruction methods used for comparison, motion is corrected for each image (both control and label).

*2.3.2. Standard image reconstruction and CBF quantification*

The standard reconstruction of ASL data was performed using direct inverse Fourier transform. Coil maps were estimated by dividing the MR image from each coil (reconstructed by inverse Fourier transform) by the root sum of squares of all images obtained from all the coils (24). The estimated motion transformations were used to compensate for motion for each control and label image. In cases where PSF deblurring was applied for the standard reconstructions, a Lucy-Richardson deconvolution (100 iterations) was performed (9). The control and label images were then subtracted and averaged to obtain a perfusion-weighted image ($\boldsymbol{x}_P$), which was converted into CBF maps (mL/100g/min) using the following equation (1).

$$\mathrm{CBF} = \frac{6000\,\lambda \times \boldsymbol{x}_P \times \exp\left(-\frac{PLD}{T_{1,b}}\right)}{2\alpha \times T_{1,b} \times \boldsymbol{x}_{M_0} \times \left(1 - \exp\left(-\frac{\tau}{T_{1,b}}\right)\right)} \qquad [5]$$

where the label duration $\tau = 1500$ ms, PLD is 1800 ms, the brain-blood partition coefficient, $\lambda = 0.9$ mL/g, the longitudinal relaxation time of blood, $T_{1,b} = 1650$ ms at 3T, and the labelling efficiency $\alpha = 0.85$ as suggested in (1); $\boldsymbol{x}_{M_0}$ is the $M_0$ calibration image corrected for the TR = 4 sec. The standard reconstructed images were then corrected for PVE using a 3DLR method implemented in MATLAB using a kernel of 5×5×5 voxels, on the ratio $\boldsymbol{x}_P/\boldsymbol{x}_{M_0}$. The FSL-FAST tool (25) was used to estimate high-resolution GM and WM PV maps from structural images, which were then transformed into the low-resolution ASL image space using FSL's *applywarp* with spline interpolation and a super resolution level of 4. In Supporting Information Figure S2, the ROIs and GM and WM PV maps obtained from the T1-MPRAGE of a participant are shown. The MOCHA method was implemented in MATLAB as summarised in Appendix A.

## 2.4. Simulations

A numerical ground truth CBF map was simulated by segmenting the T1w MR image (224×256×176 and 1.05×1.05×1.1 mm$^3$) of subject 1 into WM, GM and cerebrospinal fluid (CSF) regions using SPM. The resulting PV maps were then used to generate a CBF map through multiplying the tissue blood flows of 65 and 20 mL/100g/min by the GM PV and WM PV maps, respectively (26). Furthermore, a 1.34 mL circular WM hyper-perfused lesion with a blood flow of 100 mL/100g/min, a regional hyper-perfusion (78.9 ± 8.6 mL/100g/min) and hypo-perfusion (36.6 ± 3.8 mL/100g/min) were created to evaluate the effect of mismatches between anatomical and perfusion information on the reconstructed CBF maps. To simulate realistic high-resolution control, label and $M_0$ images, the $M_0$ and the first control k-space images of subject 1 were reconstructed in the resolution space of the T1w image using the method described in Eq. [4].

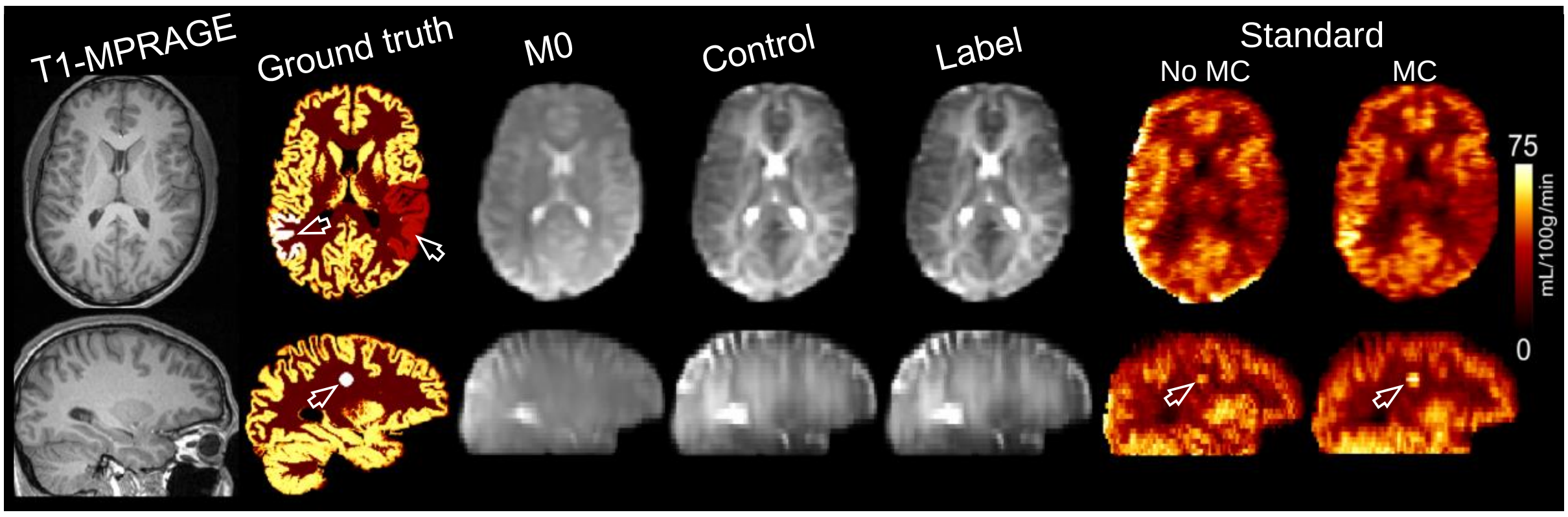

**Figure 2.** Simulated brain ASL phantom comparing the ground truth CBF map with the low-resolution CBF maps reconstructed using the standard method with (MC) and without (No MC) motion correction.

Using the simulated high-resolution CBF, $M_0$ and control images, a high-resolution label image was then created based on Eq. [5] with the default parameters. The control, label and $M_0$ images were then resampled into the resolution of ASL data (the same as our *in-vivo* data). In these simulations, 20 pairs of C-L images were considered. To simulate motion, each image was incrementally rotated, leading a maximum angular drift of 3° between the first and last C-L pair and translation of 15 mm (see Supporting Information Figure S1). The images were then downsampled to match the native resolution of our in-vivo ASL data, blurred in the partition direction using a 6 mm-FWHM Lorentzian filter, modulated by the calculated coil sensitivity maps and Fourier transformed to obtain multi-channel *k*-space dataset. Gaussian noise was added to the *k*-space data to obtain an SNR of 15 dB. Finally, the motion transformation of each C-L pair was estimated with the procedure described in *Section 2.3.1*. Figure 2 shows the high-resolution T1w and CBF images together with the simulated low-resolution $M_0$, first control and label images, and the CBF maps estimated by the standard method with and without motion correction.

## 2.5. Evaluation and parameter selection

The standard, 3DLR and MOCHA methods were evaluated for quantification of CBF in WM, cortical GM and different subcortical GM regions of the simulated and *in-vivo* datasets. The T1-MPRAGE images were parcellated into GM, WM, thalamus, caudate, putamen, pallidum and hippocampus using FreeSurfer (27). For simulations, the reconstruction methods were evaluated based on the mean CBF in different parcellated regions. For 3DLR depending on the ROI, the most appropriate of either the GM or WM PV-corrected maps was used to extract mean values. The normalized root mean square error (NRMSE), defined as:

$$NRMSE_i(\%) = 100 \times \sqrt{\frac{\sum_{j\in ROI_i}\left(x_j - x_j^{GT}\right)^2}{\sum_{j\in ROI_i}\left(x_j^{GT}\right)^2}} \qquad [6]$$

where $\boldsymbol{x}^{GT}$ is the ground truth CBF map. For the simulated data, the $\beta$ parameter of the MOCHA was optimized based on minimization of NRMSE over the whole brain, while the rest of parameters were empirically set to $\sigma = 0.15$, $\mathcal{N} = 3\times3\times3$ and $N_{it} = 100$ iterations of the SD algorithm. The same parameters were used for *in-vivo* data. The same $\beta$ was then used for *in vivo* reconstructions. For simulations, the PSF through-plane FWHM was set to 1.5 times the slice thickness to mimic an acquisition with T2 decay during the 3DGRASE echo-train. For *in-vivo* data the PSF was modelled as a Lorentzian with FWHM of 1 slice thickness. The PSF estimation was performed using autocorrelation of the residuals as described in (28). In this method, multiple C-L differences are mean-subtracted to generate voxel-level residuals. A 1-D series of residuals in the superior-inferior direction are then obtained by averaging across measurements, as well as in the anterior-posterior and left-right directions. The autocorrelation of these residuals is fitted with the autocorrelation of a Lorentzian, giving an estimate of the PSF width.

To evaluate the performance of the MOCHA method for accelerated ASL imaging, an *in-vivo* dataset was reconstructed with a retrospectively reduced number of 10 and 5 C-L pairs.

# 3. Results

## 3.1. Simulations

Figure 3 shows the simulation results of the standard, 3DLR and MOCHA reconstruction methods, (all including motion correction). As shown the standard method notably suffers from PVE and loss of details. The 3DLR method separates the GM and WM CBFs for each voxel of the standard CBF map resulting in partial recovery of estimated CBF in the GM, however, at the cost of loss of boundaries in the simulated hyper/hypo perfused regions, severe smoothing and suppression of the simulated lesion (see arrows). It is also apparent that some deep GM structures such as putamen and caudate have not been appropriately PV corrected by the 3DLR method. In contrast, MOCHA shows tissue boundaries and recovers deep GM CBF to a good extent. Due to the severe blurring introduced and low acquisition resolution simulated, uniform intensity of GM CBF across uniform regions (such as the thin cortical ribbon) cannot be achieved. Importantly, the hyper-perfused lesion, and hyper/hypo-perfused regions are well delineated, despite there being no corresponding structure on the anatomical image used for guidance.

Supporting Information Figure S3 shows the results of a similar analysis as in Figure 3 without the motion correction step, showing substantial degradation of the reconstructed maps. Importantly, no motion artefacts are apparent for MOCHA in Figure 3 with the relatively large simulated motion, despite its neglect of within-pair motion.

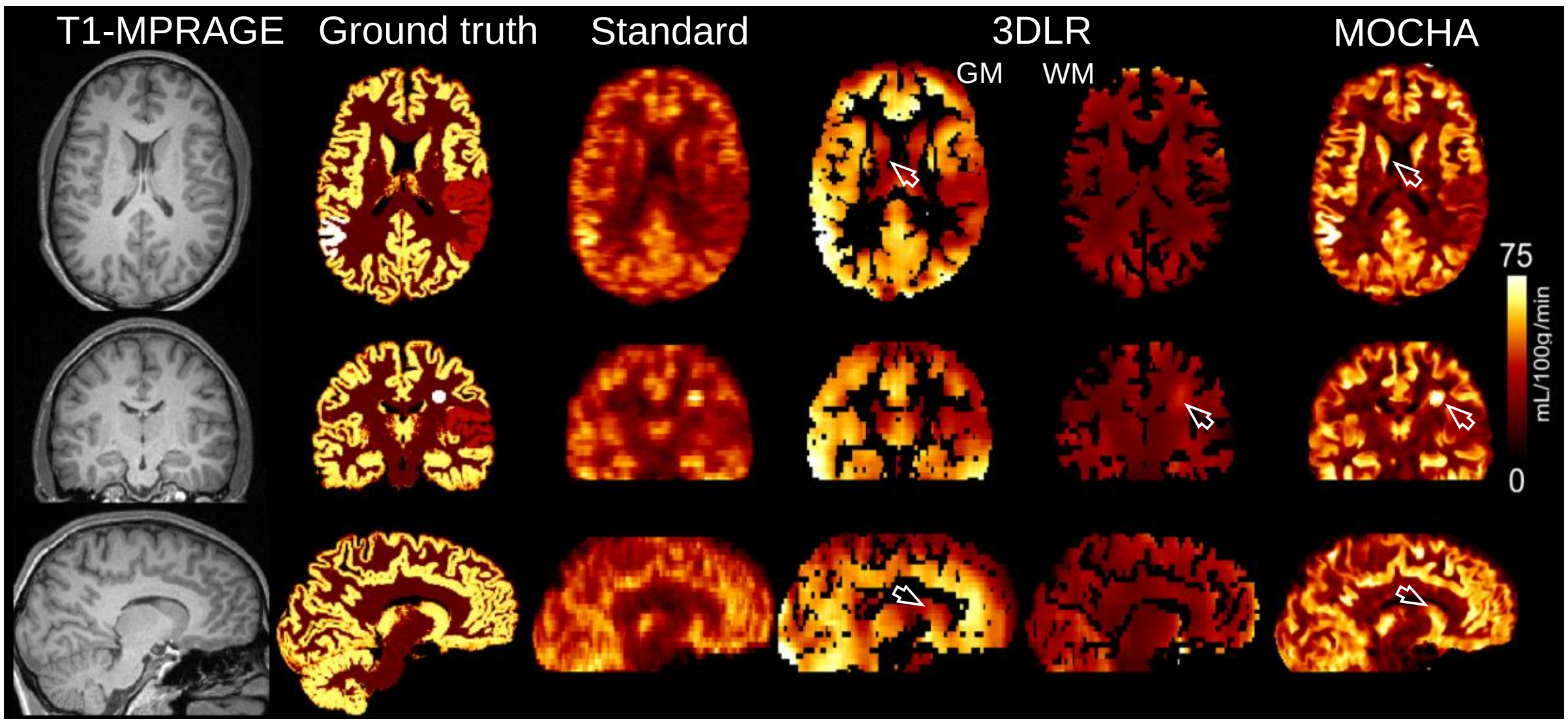

**Figure 3.** Results for reconstruction of simulated data for a motion corrected CBF map obtained from the standard method, corrected for partial volume averaging of GM and WM using the 3DLR method and reconstructed using the MOCHA high-resolution method. The arrows point to where MOCHA outperforms 3DLR in the caudate and simulated WM lesion. On the low-resolution 3DLR data the boundaries of the simulated GM lesions are also not well defined.

Figure 4 shows the performance of the methods in terms of mean and standard deviation of CBF values and NRMSE in different regions of the brain. The corresponding values (with and without motion-correction step) are summarised in Supporting Information Tables S1 and S2.

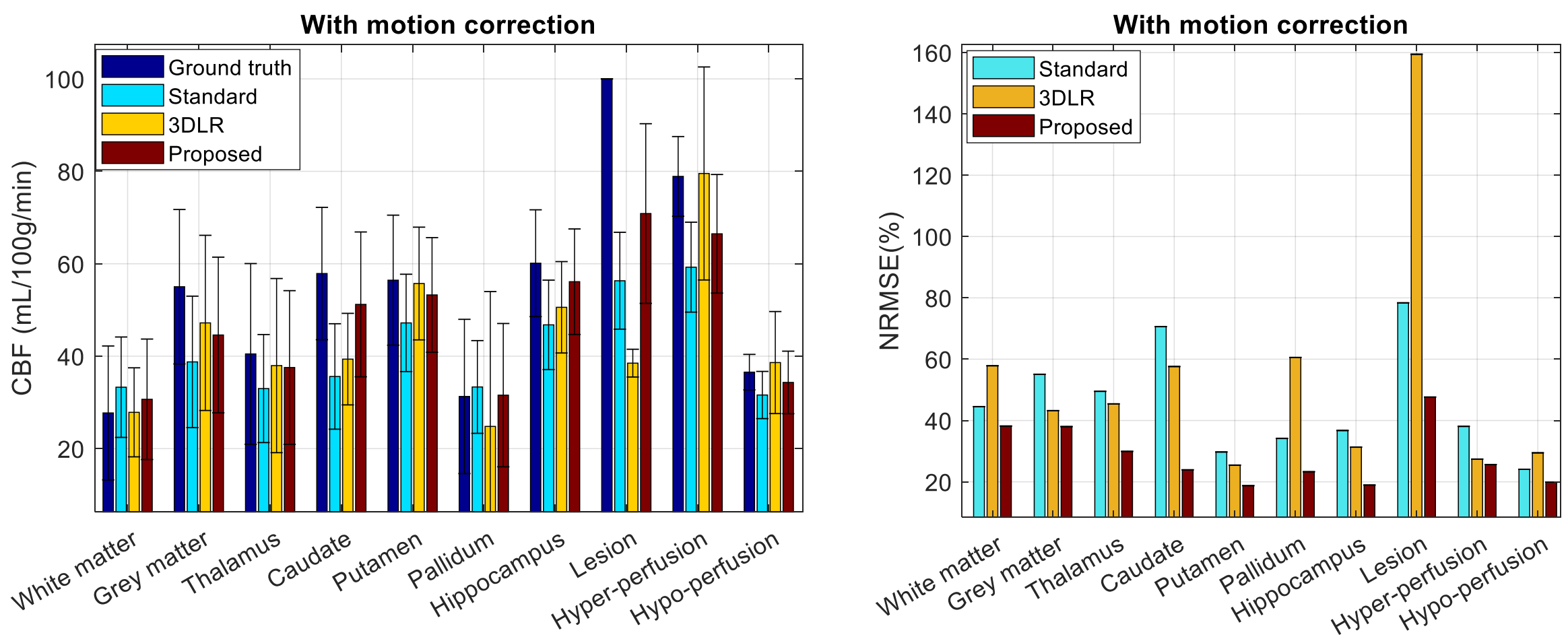

**Figure 4.** The mean and standard deviation of CBF values estimated by the studied methods in different regions of the simulated brain phantom with motion correction. Note the GM region only contains cortical GM.

In terms of mean ROI values, the standard reconstruction overestimates CBF in WM and pallidum, and underestimates CBF in all others with -44%, -25% and -14% in the WM lesion and GM hyper/hypo regions respectively (ROIs of mismatch between anatomy and perfusion). MOCHA is closer to the ground truth values than the standard reconstruction in all cases (with -29%, -16%, -6% in the mismatch ROIs). 3DLR is closer to the true values than the standard method in all regions except the pallidum and WM lesion (with -62%, +1%, +5% in the mismatch ROIs). The slightly better match to the ground truth of the 3DLR compared to MOCHA for cortical GM (-14% vs -19%) and WM (+1% vs +10%) is due to the fact that the 3DLR values reported here explicitly contain only contributions from either GM or WM 3DLR maps. MOCHA shows more accurate CBF values than 3DLR for caudate and pallidum (20% improvements), hippocampus (9% improvement) and particularly in the GM hyper-perfusion. 3DLR is slightly better than MOCHA in the putamen (-1 vs -6%), and in the GM hyper-perfusion mismatch region. Despite 3DLR showing mean ROI values closer to the ground-truth in some regions, MOCHA provided lower voxel-level NRMSE in all regions (with reductions of 112%, 2%, 10% in the mismatch ROIs vs 3DLR). NRMSE for 3DLR was higher than for the standard reconstruction in the WM, WM lesion and GM hypo-perfusion regions. Removing the motion correction step causes an increased NRMSE for all methods/regions, and a general CBF underestimation, particularly in all the anatomical/perfusion mismatch regions.

Supporting Information Figure S4 presents similar reconstructions as in Figures 3, this time with additional PSF deconvolution for the standard and 3DLR methods using the same PSF used for MOCHA. The images show improved contrast for the standard reconstruction method at the expense of noise amplification. Supporting Information Figure S5 shows that PSF deblurring slightly changes the NRMSE of standard reconstruction (on average by 4.7% reduction, variable across ROIs); for the 3DLR method there are small reductions in WM lesion, cortical and deep GM NRMSE, with a slight increase in the GM mismatch regions and WM NRMSE.

Supporting Information Figure S6 shows CBF profiles for the studied methods with respect to ground truth. As shown, PSF deblurring amplifies noise for the standard reconstruction, and slightly increases CBF for 3DLR in GM hyper-perfusion; MOCHA, which takes PSF into account in the reconstruction follows the true profiles more closely.

Supporting Information Figure S7 shows the NRMSE performance of the MOCHA as a function of the regularisation parameter $\beta$ for different regions of the simulated brain phantom. In Supporting Information Figure S8, the MOCHA reconstructions for different $\beta$ values are shown. Supporting Information Table S3 summarises the results and highlights the $\beta$ values that results in the lowest NRMSE in each region. The results show that as the $\beta$ increases the errors in the GM and especially WM reduce, however at the expense of increasing errors in the WM lesion. $\beta = 20$ was chosen for minimal errors in whole brain, and a good compromise in the simulated anatomical/perfusion mismatch regions; the same value was used for in-vivo data.

The performance of the 3DLR method was also evaluated as a function of kernel size. As shown in Supporting Information Figure S9, by increasing the kernel size the GM CBF maps are smoother and have fewer details. However, the quantitative results show that mean WM reduces very slightly for larger kernel sizes, while GM CBFs is stable; hippocampal and deep GM CBF values tend to increase very slightly (i.e. better overall PVC performance in these regions); at the same time the WM lesion's CBF notably reduces. Hence, as mentioned earlier, in this study a kernel of 5×5×5 voxels was used, which provided a balanced performance for the 3DLR method for the WM lesion and small GM structures.

### 3.2. *In-vivo* data

Figures 5 and 6 show the results of two subjects comparing different methods. Supporting Information Figures S10-S11 show similar results for another two subjects. All the *in-vivo* data was motion corrected. These results show that the standard CBF maps suffer PVE especially in partition-encoding direction. The 3DLR method results in increased GM CBF values however at the expense of some loss of details including smoothing of the apparent local high perfusion indicated by arrows in Figure 5. In comparison, the MOCHA method appears to correct for PVE while preserving local hyper-perfusions and recovering details in the partition direction (see coronal and sagittal views). Figure 7 shows the quantitative performance of the reconstruction methods in different regions of the brain, averaged over all four subjects (values in Supporting Information Table S4). Similarly to the simulation results, in these *in-vivo* data, MOCHA reduces WM CBF and increase CBF in most GM regions.

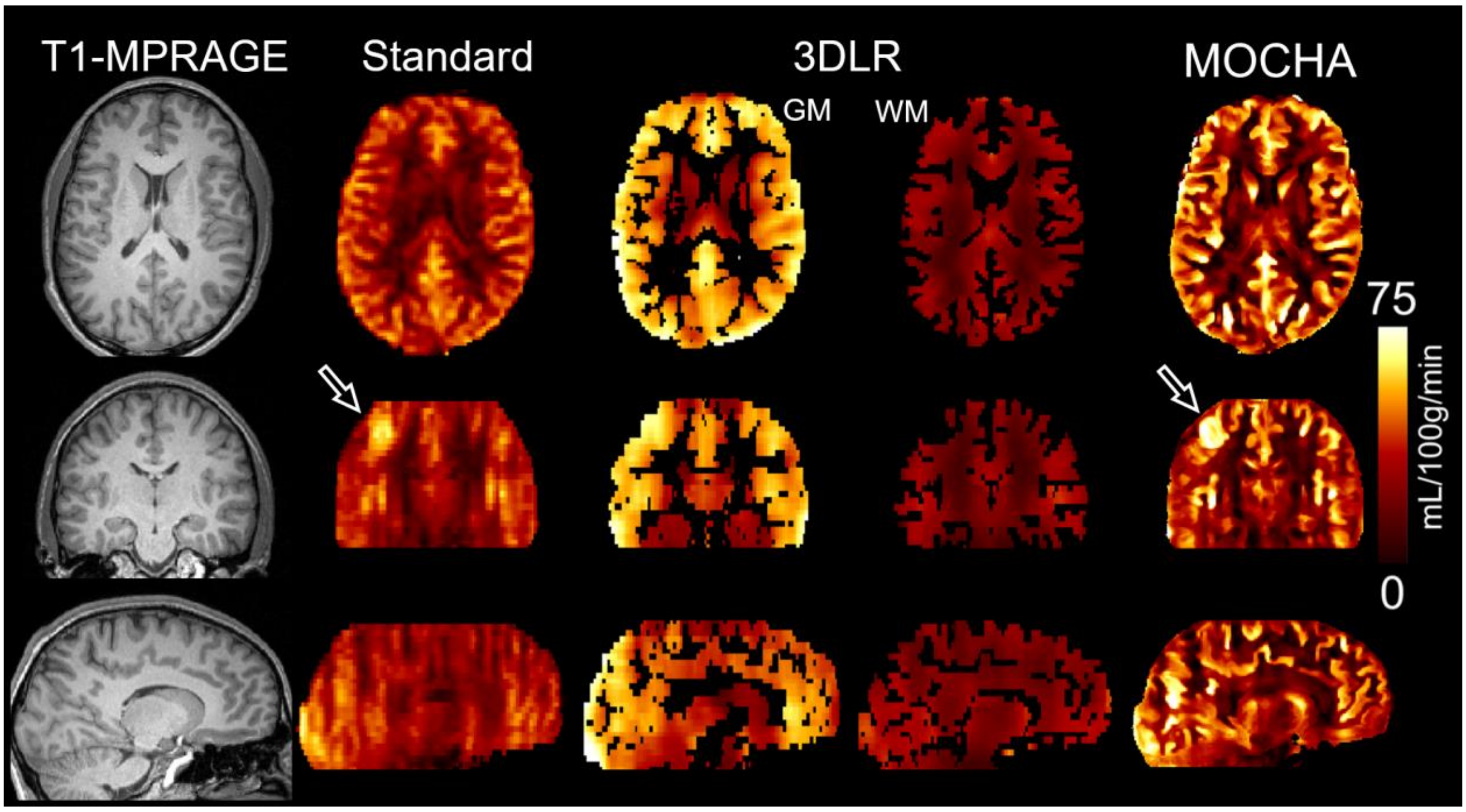

**Figure 5.** Anatomical image and CBF results for subject 1 calculated using the standard, 3DLR and MOCHA reconstruction methods. Note that data from this subject was also used for simulations. The arrows indicate an area of apparent local high perfusion.

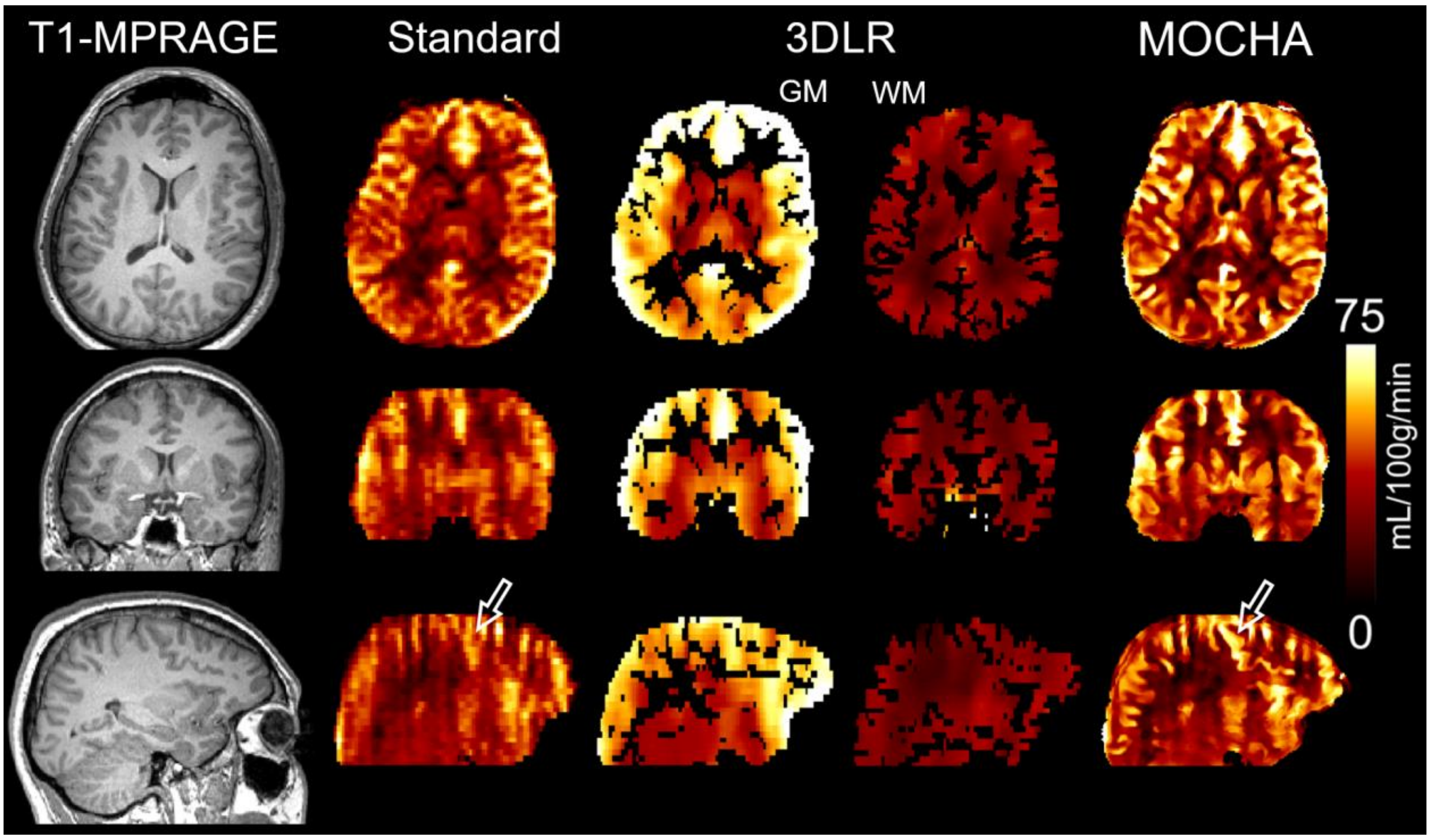

**Figure 6.** Anatomical image and CBF results for subject 2 calculated using the standard, 3DLR and MOCHA reconstruction methods.

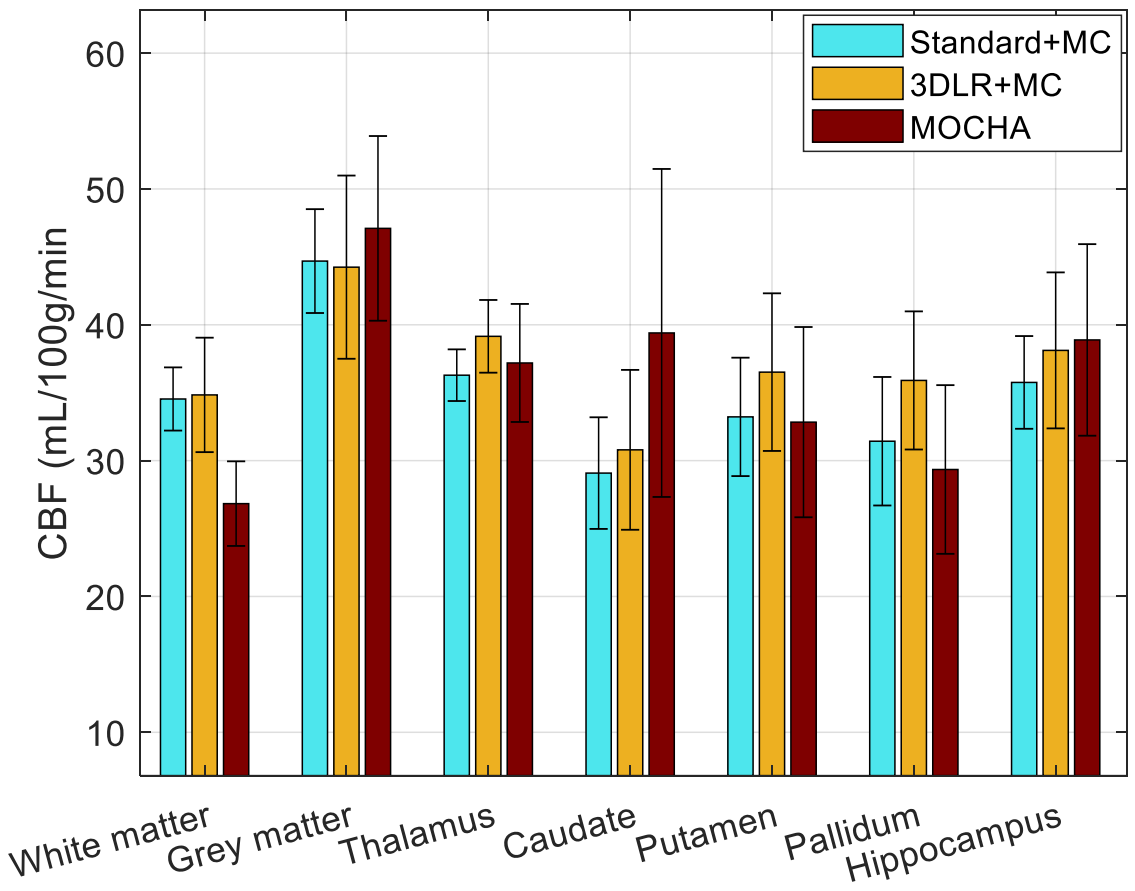

**Figure 7.** CBF results averaged over 4 *in-vivo* datasets for standard, 3DLR and MOCHA reconstruction methods. The error bars show the standard deviation of the mean CBF values calculated for each subject in each region.

Figure 8 and Supporting Information Figure S12 show the MOCHA reconstruction of the standard low-resolution data of subjects 5 and 6 compared to their corresponding higher-resolution data. As shown previously, MOCHA enhances the anatomical tissue boundaries. Most importantly, many details of the high-resolution data that are lost in the standard low-resolution reconstruction have been reliably recovered in the MOCHA reconstruction. Quantitative analysis of these results for the two volunteers are individually shown in Figure 9 and the values averaged over the two volunteers in Table S5. MOCHA not only enhances the visual appearance of the low-resolution CBF maps but also improves their quantitative accuracy towards the values found in the reference high-resolution CBF map for most ROIs. The averaged cortical GM CBF values for these volunteers were 40.2, 37.7 and 40.7 mL/100g/min for the high-resolution standard, low-resolution standard and low-resolution MOCHA reconstructions, respectively.

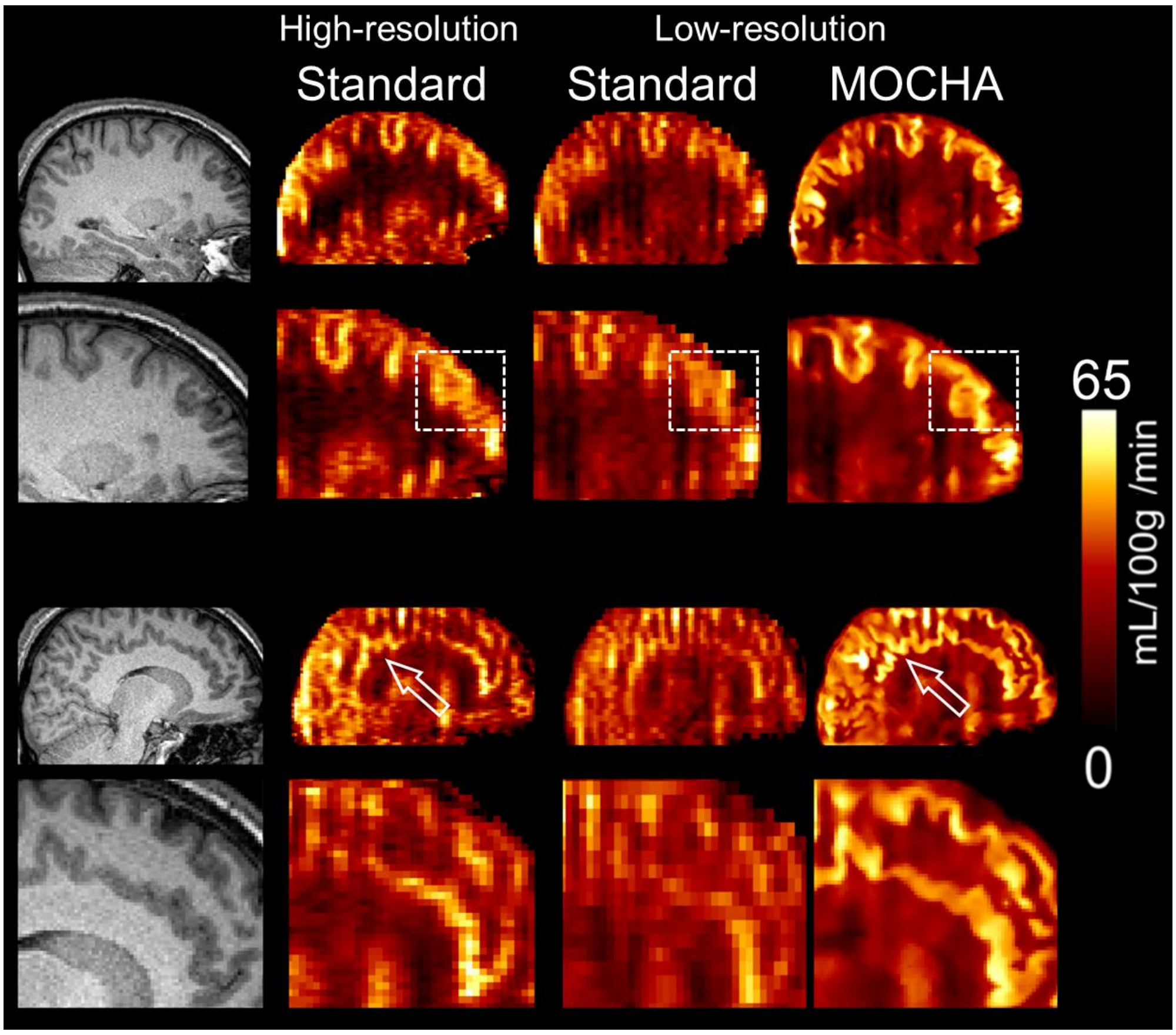

**Figure 8.** Anatomical image and CBF maps from standard-resolution (4×4×4 mm$^3$; 5 min 40 s acquisition; standard and MOCHA reconstructions; right) and doubled resolution in the inferior-superior direction ('high-resolution'; 4×4×2 mm$^3$; 22 min acquisition; standard reconstruction; left) for subject 5. Note that due to the sequential nature of the acquisitions, there might be physiological differences between low-resolution and high-resolution CBF maps.

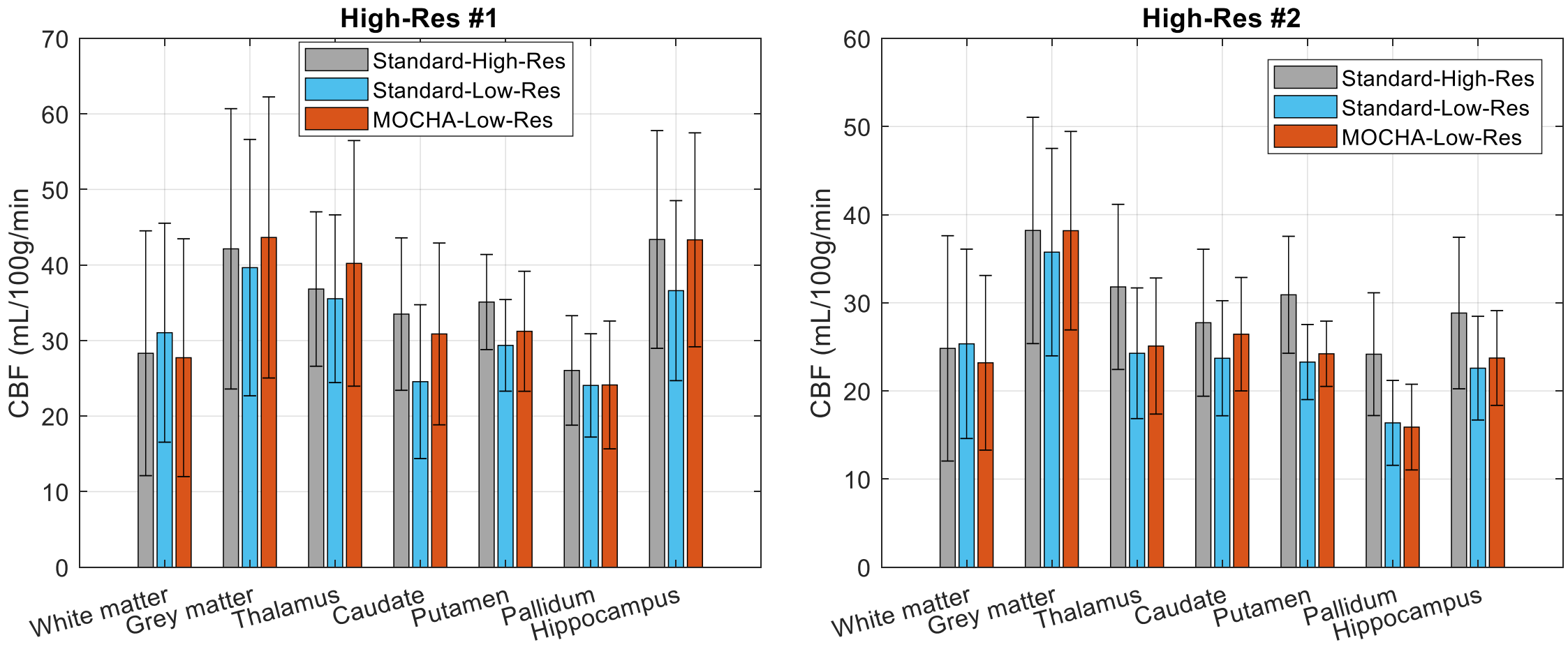

**Figure 9.** ROI averaged CBF results for the low-resolution standard and MOCHA reconstructions of the standard resolution acquisition (2mm slice thickness), compared with the two high-resolution ASL in-vivo scans (2mm slice thickness, left; and 1.33mm slice thickness, right, respectively). The error bars show the standard deviations over each region.

The performance of MOCHA was further evaluated for acquisitions with a lower number of C-L pairs (or repeats), which would entail reduced scan time and correspondingly reduced SNR. For this purpose, a dataset has retrospectively been reduced to 10 and 5 C-L pairs out of 20 pairs, equivalent to SNR reductions of 1.4 and 2 respectively. Figure 10 compares the reconstruction results of the standard and MOCHA methods. As shown, for a lower number of C-L pairs the standard CBF map appears slightly noisier compared to the reference 20-pairs image, whereas MOCHA shows more consistent maps. Supporting Information Table S6 summarises the quantitative performance of the methods. The results show a slight GM CBF decrease with the increase of C-L pair used, which could be potentially attributed to a physiological decrease of CBF during the 6-min acquisition. The expected acquisition times for 5-pair and 10-pair acquisitions are, including dummy scans and M0 data collection, 100 s and 180 s compared to 340 s of the reference 20-pair scan.

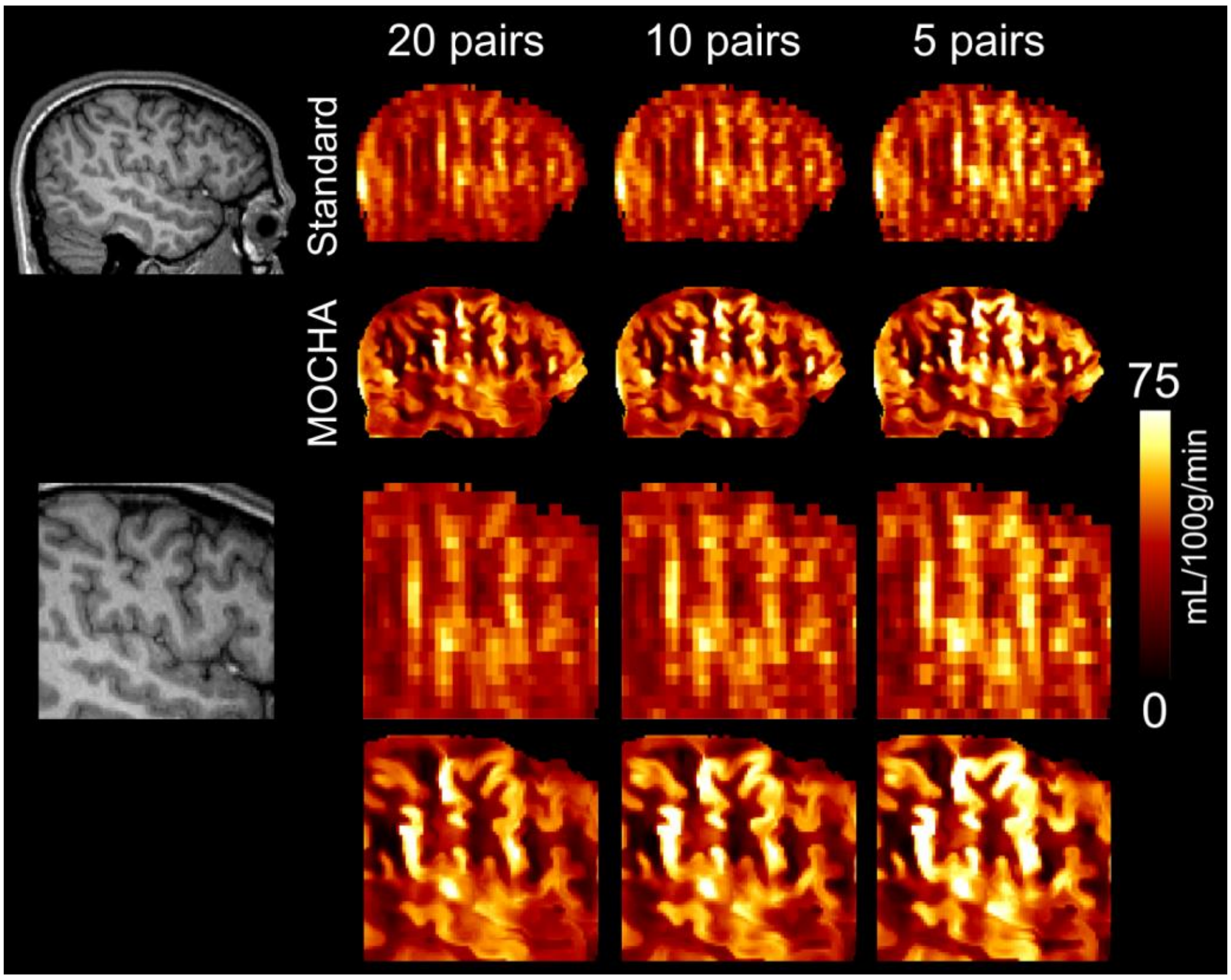

**Figure 10.** CBF results for subject 4 calculated using standard and MOCHA reconstruction methods using different numbers of C-L pairs (i.e. 1-20, 1-10 and 1-5).

The performance of MOCHA was also evaluated for undersampled ASL scans. For this purpose, the k-space data of subject 4 was retrospectively undersampled at two levels (acceleration factor R of 2 and 4) in the phase-encoding (anterior-posterior) direction. The performance of MOCHA was then compared with the standard CBF maps reconstructed using sensitivity encoding (SENSE) and SENSE with total variation regularisation (TV) (29). As shown in Supporting Information Figure S13, MOCHA reduces noise and undersampling artefacts and maintains an image quality similar to fully sampled data, demonstrating good potential for undersampled acquisitions with highly accelerated acquisition times.

# 4. Discussion

In this study, the proposed MOCHA reconstruction framework was compared to the 3DLR PVC method.

The 3DLR method separates the GM and WM signals within each voxel of the standard low-resolution CBF maps by solving a system of equations in which the GM and WM PV fractions are known coefficient values. This method assumes all voxels in the neighbourhood (kernel) of a given voxel have the same GM and WM CBF values hence the system is uniquely solved by a least squares estimator, however at the expense of smoothing image details, as shown here and by others. Recently, a Bayesian approach was proposed to solve the underdetermined system by employing the kinetic model of the GM-WM signals in multi-PLD ASL acquisitions together with a prior modelling the spatial correlation of kinetic parameters (6). Since for our study, the data were acquired with single PLD and perfusion was quantified using Eq. 5, recommended by (1), rather than using a kinetic curve fitting as in (6), the Bayesian method was not included to avoid any discrepancy caused by the perfusion estimation method. However, Oliver *et al* compared the 3DLR (3×3×3 kernel) and Bayesian methods in 6 healthy controls (17). Despite comparable mean GM-CBF values, the Bayesian method retained structural details at the expense of increased sensitivity to noise. Recently, Zhao *et al* compared a 2DLR (3×3 kernel) with the Bayesian method using comprehensive simulations and *in vivo* data (30) and similar results were observed.

In contrast to these two methods, MOCHA aims to reconstruct directly a high-resolution CBF map corrected for different sources of PVE, such as large voxel sizes, PSF and motion blurring. MOCHA employs all C-L pairs to reconstruct a single perfusion-weighted image, such that the averaging is performed during reconstruction rather than after reconstruction of the individual C-L pairs. The same idea has recently been used by (31) to explore temporal redundancy and spatial similarity of the C-L pairs for ASL reconstruction. As expected, in simulations both 3DLR and MOCHA produced averaged GM/WM CBF values closer to ground truth than the standard reconstruction. MOCHA provided sharper anatomical boundaries, whilst 3DLR showed increased blurring. The 3DLR mean values were slightly closer to the ground truth in cortical GM, and hyper-perfused GM, whereas MOCHA performed a lot better in the WM hyper-perfusion region. Furthermore, MOCHA provided the lowest NRMSE in all regions analysed.

MOCHA relies on anatomical images for regularisation of high-resolution CBF maps. Whilst this improves the quality of the reconstructed CBF maps, we are aware that some functional features (i.e. geometry of flow territories, vascular artefacts) influencing CBF are not captured by anatomy (i.e. GM/WM/CSF PV fractions or tissue appearance on T1w images); hence any method (including LR) relying on anatomical information for PVE correction could lead to partly biased results, and a high-resolution perfusion signal cannot be completely recovered solely by using anatomical information. At the same time we have shown that by combining low-resolution perfusion data and high-resolution anatomical information MOCHA does go some way towards correcting functional maps for PVE and blurring and thus improving their spatial and quantitative accuracy. We have considered a number of scenarios to demonstrate this by simulating anatomy/perfusion mismatches, i.e. hyper/hypo perfusion in GM and WM regions with no corresponding structural abnormalities. Whilst MOCHA's quantitative accuracy varies depending on the region, it always offers an improvement compared to the standard reconstruction, and in all cases provides an improved preservation of boundaries and the lowest voxel-level errors (NRMSE) compared to the ground truth. In these situations, 3DLR's quantitative performance is also variable and, not being able to rely on PV information, it is inherently affected by large blurring, the extent of which depends on the chosen kernel size.

We have also used in-vivo datasets acquired with high-resolution protocols as high-quality references to validate the MOCHA reconstructions obtained from standard low-resolution (4 $mm^3$) 6 min acquisition time datasets. To obtain these references we doubled and tripled the slice resolution, requiring long acquisitions of 22 min and 49 min respectively for full k-space sampling with equivalent SNR. It was apparent that many of the details in the long-acquisitions/high resolution CBF maps are in fact well reproduced in the MOCHA images reconstructed from low-resolution data. This suggests an effective resolution improvement which is beyond purely visual improvement.

Finally, we have provided evidence that MOCHA-reconstructed CBF maps are robust to a severe reduction in the number of C-L pairs (averages) collected, with reductions tested up to a factor of 4, and k-space undersampling, also tested up to a speed-up factor of 4. While there are clear advantages, MOCHA nonetheless has some limitations. Inclusion of PSF and downsampling and using an anatomical prior leads to only partial recovery of the lost high-frequency information. In (32), a similar idea of transferring the high-frequency information from structural MR to low-resolution emission tomography data has been proposed without segmentation of the MR image. The PSF was assumed to be shift-invariant and motion-independent. Following (33), the blurring $\boldsymbol{B}$ was therefore used as the frontend operator in Eq. [1], which allows the motion transformation and downsampling operators to be merged into one single spatial transformation, reducing the computational burden of the model.

As tissue boundaries of the anatomical data influence the MOCHA reconstruction of the perfusion images, any motion left unaccounted for, as well as any distortions or misregistrations affecting the alignment of the structural and perfusion data, can all negatively affect the accuracy of the reconstruction. Whilst a number of steps were taken to minimise these effects, further improvements are possible and will be undertaken. The current MOCHA implementation only takes into account motion occurring between C-L pairs and neglects motion during each acquisition/pair. Whilst for the continuous motion in our simulations and our healthy volunteers the current inter-frame motion correction appears to be sufficient, more complex motion patterns may occur, especially for non-cooperative patients. A possible solution to address frame-by-frame motion is to reconstruct a motion-corrected control image from all control data and likewise for all label data, and then to perform a post-reconstruction subtraction. Additionally, it is possible to identify motion-corrupted 'outlier' pairs (e.g. using ENABLE (34)) and remove them from the analysis. In our data with limited PSF blurring, the $M_0$ to T1w rigid-body registration produced satisfactory registration. However we are aware that in general, for 3D GRASE ASL, which suffers from T2-blurring, registering CBF maps with GM PV maps, as in (35), has been shown to be more reliable; this method is therefore recommended for a more general application of MOCHA.

In this paper, susceptibility-induced geometric distortions were not included in our forward model. Instead they were minimised by the 2-fold segmentation in the phase encoding (anterior-posterior) direction. However, small residual localised spatial mismatches between anatomical and perfusion images can remain. Future work will include estimation and correction of the spatial distortions, for example using reversed gradient (blip up/down) acquisitions, further reducing this potential source of error.

All these various misalignment errors discussed above can cause some local or global CBF errors. However their magnitude also depends on the strength of the regularization ($\beta$) and the shape of the Gaussian similarity coefficients ($\sigma$). In this work, $\sigma, \mathcal{N}$(neighbourhood size) and the number of iterations were chosen heuristically. Larger values of $\sigma$ reduce the impact of MR information, as the resulting weights will tend to be more uniform. We have found $\sigma$ values in the range of 0.1-0.3 result in appropriate weighting of the structural information from T1w images. $\mathcal{N}$ was set to 3×3×3, as in our CPU-based implementation, larger neighbourhoods are memory demanding, and based on our previous experience larger neighbourhoods do not notably lead to improved regularisation. We used a large number of iterations to ensure that the steepest descent algorithm (which improves upon gradient descent by step size optimisation) converges to at least a fixed point. In our experience, the most important hyperparameter is $\beta$, which was optimised.

For the main results, we compared MOCHA to 3DLR with a 5×5×5 voxel kernel and no PSF modelling. The kernel size of the LR method impacts on its performance in terms of robustness to noise (30), geometric distortion (36) and of repeatability between scans (37). This was chosen based on our tests with kernels ranging from 3×3×3 to 9×9×9 voxels. Taking into account the PSF for 3DLR only provided a small reduction in NRMSE in GM ROIs and the WM lesion but not in the GM hypo/hyper perfusion regions; therefore for the main results PSF modelling was not included, which is consistent with the most common usage of 3DLR in the existing literature. Admittedly the 3DLR performance in some deep GM regions could have been improved with a different PV estimation method, for instance the recently developed tissue probability maps with better subcortical performance (38). However this would not have improved 3DLR results in areas of anatomical/functional mismatches and highlights the dependence of the 3DLR on PV tissue fraction and its estimation method; one advantage of MOCHA is that it does not even require PV maps.

Our method is computationally intensive due to the inclusion of motion, spatial mapping and PSF operators in the forward model. The computation time for 1 and for 100 iterations in MATLAB R2017a (running on a 20-core Intel Xeon 3.10-GHz workstation, for a dataset of 20 averages) was ~1 minute and 1.5 hours, respectively. MOCHA's objective function is convex and continuously differentiable, hence the steepest descent algorithm guarantees convergence to the global minimizer, irrespective of the initial estimate.

To our knowledge MOCHA is the first fully model-based high-resolution reconstruction method for ASL data. Our results show a good performance of MOCHA in simulations including areas of anatomical/perfusion mismatch. In-vivo data demonstrated that MOCHA can reliably reconstruct high-resolution CBF maps from standard low-resolution datasets, featuring many of the details observed in the higher resolution datasets (which require impractically long acquisition times). The robustness of the reconstruction to short acquisitions and/or undersampling was also demonstrated. The actual clinical benefits of MOCHA will be evaluated in collaboration with radiologists, using patient data presented with and without MOCHA reconstruction.

# 5. Conclusion

Simulation and *in-vivo* data results demonstrate that the proposed direct high-resolution CBF map reconstruction method effectively corrects motion and PVE. MOCHA is advantageous in preservation of structural details and hypo/hyper-perfused regions. The MOCHA framework has potential to improve the diagnostic confidence and applicability of current ASL protocols in clinical practice.

# Acknowledgements

This work is supported by the Engineering and Physical Sciences Research Council (EPSRC) under grant EP/M020142/1 and the Wellcome EPSRC Centre for Medical Engineering at King's College London (WT 203148/Z/16/Z). CJM is supported by MRC grant MR/N013042/1. According to EPSRC's policy, all data supporting this study will be openly available at https://doi.org/10.5281/zenodo.1470571

# Appendix A

**The steepest descent (SD) algorithm**

- Initialize parameters: $\boldsymbol{x}_p^{(0)} = \boldsymbol{x}_{M_0}^{(0)} = \boldsymbol{d}\phi_p = \boldsymbol{0},\ \sigma, \beta, \boldsymbol{W} = \boldsymbol{I}$
- Calculate the proximity and similarity coefficients, $\boldsymbol{\xi}, \boldsymbol{\omega}$, as defined in Eqs. [3,4].
- Define the operators: $(\cdot)^H$ as conjugate transpose, and $\boldsymbol{d}R$, derivative of Eq. [2], and $F$ as follows:

$$\boldsymbol{d}R(\boldsymbol{z}, \boldsymbol{\omega}) = \beta \sum_{b \in \mathcal{N}_j} \omega_{jb} \xi_{jb} \left(z_j - z_b\right), j = 1, \dots, N_h$$

$$F(\boldsymbol{z}, \boldsymbol{\omega}) = \left(\boldsymbol{E}\boldsymbol{T}_{M_0}\boldsymbol{B}\right)^H \boldsymbol{E}\boldsymbol{T}_{M_0}\boldsymbol{B}\boldsymbol{z} + \boldsymbol{d}R(\boldsymbol{z}, \boldsymbol{\omega})$$

**For** $k = 0, \dots, N_{it}$

**1.** Calculate the gradient of the cost functions in Eqs. [1, 5]:

**For** $i = 1, \dots, N_p$

$$\boldsymbol{d}\phi_p \leftarrow \boldsymbol{d}\phi_p + \frac{1}{N_p} \left(\boldsymbol{E}\boldsymbol{T}_i\boldsymbol{B}\right)^H \left(\boldsymbol{E}\boldsymbol{T}_i\boldsymbol{B}\boldsymbol{x}_p^{(k)} - \boldsymbol{d}_i\right)$$

**End**

$$\boldsymbol{d}\phi_p \leftarrow \boldsymbol{d}\phi_p + \boldsymbol{d}R\left(\boldsymbol{x}_p^{(k)}, \boldsymbol{\omega}\right)$$

$$\boldsymbol{d}\phi_{M_0} = \left(\boldsymbol{E}\boldsymbol{T}_{M_0}\boldsymbol{B}\right)^H \left(\boldsymbol{E}\boldsymbol{T}_{M_0}\boldsymbol{B}\boldsymbol{x}_{M_0}^{(k)} - \boldsymbol{s}_{M_0}\right) + \boldsymbol{d}R\left(\boldsymbol{x}_{M_0}^{(k)}, \boldsymbol{\omega}\right)$$

**2.** Calculate complex-valued optimal step sizes:

$$\alpha_p = \frac{\left(\boldsymbol{d}\phi_p\right)^H \boldsymbol{d}\phi_p}{\left(\boldsymbol{d}\phi_p\right)^H F\left(\boldsymbol{d}\phi_p, \boldsymbol{\omega}\right)}, \qquad \alpha_{M_0} = \frac{\left(\boldsymbol{d}\phi_{M_0}\right)^H \boldsymbol{d}\phi_{M_0}}{\left(\boldsymbol{d}\phi_{M_0}\right)^H F\left(\boldsymbol{d}\phi_{M_0}, \boldsymbol{\omega}\right)}$$

**3.** Update images:

$$\boldsymbol{x}_p^{(k+1)} \leftarrow \boldsymbol{x}_p^{(k)} - \alpha_p \boldsymbol{d}\phi_p$$

$$\boldsymbol{x}_{M_0}^{(k+1)} \leftarrow \boldsymbol{x}_{M_0}^{(k)} - \alpha_{M_0} \boldsymbol{d}\phi_{M_0}$$

**End**

# Supporting Information

**Table S1.** Quantitative performance of the standard, 3DLR and MOCHA methods in terms of cerebral blood flow (mean ± standard deviation) in different regions of the simulated brain phantom with and without motion correction.

**Table S2**. NRMSE (%) of the studied methods with and without motion correction in different regions of the simulated brain phantom.

**Table S3**. The NRMSE performance of MOCHA in different regions of the simulated brain phantom as a function the regularisation parameter $\beta$.

**Table S4**. Mean and SD of CBF values averaged over 4 healthy subjects.

**Table S5**. Mean and SD of CBF values averaged over the 2 high-resolution healthy subjects.

**Table S6**. Mean and SD values of CBF maps for a subject calculated for different control-label pairs.

**Figure S1.** Simulated motion (translation and rotation) in our simulation dataset.

**Figure S2.** The ROIs and GM partial volume estimates obtained from the parcellation of the T1-MPRAGE MR image using the Freesurfer and FSL (FAST and *applywarp*) software. The arrow points to the pallidum that has been erroneously identified as WM.

**Figure S3.** Same as Figure 3, except for the omission of the motion correction step.

**Figure S4.** Same as Figure 3, except for the addition of PSF deblurring for the standard and 3DLR methods. NB: the motion correction step is included for all methods.

**Figure S5.** Impact of PSF deblurring on NRMSE performance of the standard and 3DLR methods for simulations.

**Figure S6.** CBF profiles of the studied reconstruction methods through simulated WM lesion and GM hyper-perfusion

**Figure S7.** The NRMSE performance of MOCHA in different regions of the simulated brain phantom as a function of the regularization parameter $\beta$.

**Figure S8.** The reconstruction results of the MOCHA method as a function of regularisation parameter.

**Figure S9.** The impact of kernel size on the qualitative (top) and quantitative (bottom) performance of the 3DLR method in comparison with the standard method and the ground truth simulated brain phantom.

**Figure S10.** CBF results for the subject 3 calculated using the standard, 3DLR and MOCHA reconstruction methods. The arrows point to where there are most notable difference between MOCHA and standard reconstruction methods.

**Figure S11.** CBF results for the subject 5 calculated using the standard, 3DLR and MOCHA reconstruction methods. The arrows point to some regions where there are notable difference between MOCHA and standard reconstruction methods.

**Figure S12.** Anatomical image and CBF maps from standard-resolution acquisition (4×4×4 $mm^3$; 5 min 40 sec acquisition) standard and MOCHA reconstructions; right) and tripled resolution in the slice direction ('high-resolution', 4x4x1.33 $mm^3$; 49 min acquisition; standard reconstruction only; left) datasets for subject 6.

**Figure S13.** CBF results of the subject 4 calculated using the standard and MOCHA reconstruction methods for different undersampling factors (R). The arrow shows an undersampling artefact. Standard and SENSE reconstructions show increased noise as R increases. TV-SENSE also shows visible changes between R=2 and R-4. MOCHA shows the highest visual consistency between reconstructions at R=1, R=2 and R=4.